\documentclass[letterpaper,10pt,twoside]{article}
\usepackage{amssymb}
\usepackage{amsmath}
\usepackage{latexsym}
\usepackage{verbatim}
\usepackage[active]{srcltx}
\usepackage{epsfig}
\usepackage{stmaryrd}
\usepackage{fancyhdr}
\usepackage{pstricks,pst-node}
\usepackage{rotating,graphicx}
\usepackage[english]{babel}

\newlength{\enviropost}
\newcommand{\be}{\begin{equation}}
\newcommand{\ee}{\end{equation}}
\newcommand{\ble}[1]{\begin{equation} \label{#1}}
\newcommand{\bae}{\begin{eqnarray}}
\newcommand{\eae}{\end{eqnarray}}
\newcommand{\fle}[2]%
{\vspace{1.5ex}
\be
\label{#1}
\mbox{%
\setlength{\fboxsep}{3ex}%
\framebox{$\dss #2 $}}
\ee} 
\newcommand{\flec}[2]%
{\vspace{1.5ex}
\be
\label{#1}
\mbox{%
\setlength{\fboxsep}{3ex}%
\framebox{$\dss #2 $}}
\, \, \, ,
\ee} 
\newcommand{\flep}[2]%
{\vspace{1.5ex}
\be
\label{#1}
\mbox{%
\setlength{\fboxsep}{3ex}%
\framebox{$\dss #2 $}}
\, \, \, .
\ee}

\newtheorem{state}{S$\! \!$}
\newtheorem{defin}{D$\! \!$}
\newtheorem{exatitle}{Example}
\newtheorem{problemdef}{Problem}
\newtheorem{soldef}{Solution}
{\textsc{Definition:} \ }% 
{\vspace{\enviropost} \\}
{\noindent \textsc{Proof}:\ }% 
{\hfill $\blacksquare$ \vspace{\enviropost} \\}
{\begin{soldef} 
\label{#2:Sol} 
#1
\marginpar{(\textbf{\pageref{#2:Pro}})}
\end{soldef}}% 
{\hfill $\Box$ \vspace{.5\enviropost} \\}
{\begin{exatitle} \label{#2} #1 \end{exatitle}}% 
{\hfill $\Box$ \vspace{\enviropost} \\}
{\begin{exatitle} \label{#2} #1 \end{exatitle}}%
{\hfill $\Box$ \\}
{\begin{problemdef} 
\label{#2:Pro} 
#1 
\marginpar{(\textbf{\pageref{#2:Sol}})}
\end{problemdef}}% 
{\hfill \vspace{.5\enviropost} \\}
{\noindent {\bf Aside:} \small}% 
{\hfill \rule[-3mm]{0mm}{0mm}$\Diamond$\\}
{%
\vspace{3mm} 
\begin{center}
\begin{minipage}{.8\textwidth} 
\begin{state} 
\label{#1}%
}% 
{%
\end{state}
\end{minipage}
\end{center}
\vspace{3mm}
}
{%
\vspace{3mm} 
\begin{center}
\begin{minipage}{.8\textwidth} 
\begin{defin} 
\label{#1}%
}% 
{%
\end{defin}
\end{minipage}
\end{center}
\vspace{3mm}
}

\newcommand{\dss}{\displaystyle}

\newcommand{\eg}{\hbox{\em e.g.{}}}
\newcommand{\etc}{\hbox{\em etc.{}}}
\newcommand{\ie}{\hbox{\em i.e.{}}}
\newcommand{\wrt}{\hbox{w.r.t.{}}}

\newcommand{\rhs}{\hbox{r.h.s.{}}}

\newcommand{\calM}{\mathcal{M}}

\newcommand{\calO}{\mathcal{O}}

\newcommand{\papertitle}{%
Operational Geometry on de Sitter Spacetime
}
\newcommand{\runningtitle}{%
Operational Geometry on de Sitter Spacetime
}
\newcommand{\paperauthor}{%
P.{} Aguilar, Y.{} Bonder, C.{} Chryssomalakos, and D.{} Sudarsky
}
\begin{document}
%%%%%%%%%%%%%%%%%%%%%%%%%%%%%%%%%%%%%%%%%%%%%%%%%%%%%%%%%%%%%%
\begin{titlepage}
\vspace*{-1cm}
\begin{flushright}
\textsf{}
%\\
%\textsf{ICN-UNAM-yy/pp}
\\
\mbox{}
\\
\textsf{\today}
\\[3cm]
\end{flushright}
\renewcommand{\thefootnote}{\fnsymbol{footnote}}
\begin{LARGE}
\bfseries{\sffamily \papertitle}
\end{LARGE}

\noindent \rule{\textwidth}{.6mm}

\vspace*{1.6cm}

\noindent \begin{large}%
\textsf{\bfseries%
\paperauthor
}
\end{large}

%\vspace*{.1cm}

\phantom{XX}
\begin{minipage}{.8\textwidth}
\begin{it}
\noindent Instituto de Ciencias Nucleares \\
Universidad Nacional Aut\'onoma de M\'exico\\
Apartado Postal 70-543, 04510 M\'exico, D.F., M\'EXICO \\[3mm]
\end{it}
\texttt{pedro.aguilar,yuri.bonder,chryss,sudarsky@nucleares.unam.mx}\\
\texttt{ 
\phantom{X}}
\end{minipage}
\\

\textsc{\large Abstract: }
Traditional geometry employs idealized concepts like that of a point or a curve,
the operational definition of which relies on the availability of classical
point particles as probes. Real, physical objects are quantum in
nature though, leading us to consider the implications of using
realistic probes in defining an effective spacetime geometry. As an example, we
consider de Sitter spacetime and employ the centroid of various composite probes
to obtain its effective sectional curvature, which is found to depend on the
probe's internal energy, spatial extension, and spin. Possible refinements of
our approach are pointed out and remarks are made on the relevance
of our results to the quest for a quantum theory of gravity. 
\end{titlepage}

\setcounter{footnote}{0}
\renewcommand{\thefootnote}{\arabic{footnote}}
\setcounter{page}{2}
\noindent \rule{\textwidth}{.5mm}

\tableofcontents

\section{Introduction}
\label{Intro}
%%%%%%%%%%%%%%%%%%%%%%%%%%%%%%%%%%%%%%%%%%%%%%%%%%%%%%%%%%%%%%
%%%%%%%%%%%%%%%%%%%%%%%%%%%%%%%%%%%%%%%%%%%%%%%%%%%%%%%%%%%%%%
Classical geometry, which lies at the foundations of our physical theories, is
based on concepts like points, curves, tangent vectors, \etc, the operational
definition\footnote{By \emph{operational} definition we mean one involving
(possibly \emph{gedanken}) experiments, the outcomes of which determine the
values of the geometric quantities in question.} of which relies on the
availability of classical point particles, with definite position and velocity.
However, quantum theory indicates that there are no objects to be found in nature that can
be considered as having definite positions and velocities. What can be hoped
for, at least under favorable conditions, is that some effective geometry could be
read off ``experimentally'', which might depend not only on the geometry of 
the underlying manifold, but also on the particular characteristics of the experiments 
conducted and the probes used. 

Our motivating goal of defining geometry operationally, taking into account
that realistic probes are quantum objects, is certainly a formidable task ---
what we hope to accomplish here is to take some first steps in that direction,
hoping to elucidate features that would persist in a more exhaustive analysis (related studies have appeared, in a different context, in Ref. ~\cite{Pul_Gam:07}).
To keep things tractable,  we assume that a classical underlying geometry is
given. This represents already a substantial simplification, as the full scale
problem, as encountered in, \eg, loop quantum gravity, spin
foams, dynamical triangulations, the poset program, \etc{}, involves a quantum
state of the gravitational degrees of freedom, from which the
effective classical geometry is to be extracted. The point is that we regard the
operational definition as providing physical meaning to the concepts
involved, allowing the interpretation of the results of a physical theory.
Without such operational definitions, even in the presence of a well defined
mathematical structure, with known solutions, one faces the problem
of interpretation for which one has no well established criteria and, thus, must
rely on methods based mostly on intuition and the plausibility of the results.
Therefore, although the present work has no direct connection with
particular theories of quantum gravity, we believe that the lessons drawn
from our preliminary investigations may have  an impact
on the way such theories recover the classical regime in appropriate limits. 

The full scale problem alluded to above has two main complicating features,
the backreaction of the probe, and its quantum nature. The latter shows up in the fact 
that probe wavefunctions have support over extended regions of spacetime, so that  a kind
of quantum average of the underlying classical geometry is performed when using
the probe.
It is precisely a classical analogue of this that we aim at, by employing
extended classical probes, and pondering on the kind of effective geometry
that can be read off by using them. The first step in this direction involves
assigning an effective position to an extended classical object in a curved
spacetime. Of the many available prescriptions, all of which generalize the
Newtonian center-of-mass concept, the most familiar is perhaps that of
Dixon and Beiglb\"ock \cite{CM}, which gives rise to a covariant center-of-mass worldline.
In spite of this virtue, the practical difficulty of its calculation renders it
unsuitable for our purposes, and we use instead the \emph{centroid}, which,
roughly speaking, is an energy-weighted (and observer-dependent) average position.  
We have found,
in our preliminary analysis, that the generic characteristics of our results are
independent of the particular choice of effective position and that the
complications of working with the center-of-mass do not add any valuable
insights at the level of the studies undertaken in the present paper. 

In general, the centroid worldline, just as that of the center-of-mass, fails to
be a geodesic of the background metric, which we take here to be that of de
Sitter. The gist of our method resides in
declaring such a curve to be a geodesic of an operationally defined effective
geometry.
That is, if the above extended objects are the only available probes, the
operational definition of spacetime geometry needs to be based on those
objects' behavior. An effective sectional curvature is then extracted, 
\emph{a la} Jacobi, by looking
at the relative acceleration of neighboring centroid worldlines, and its
dependence on characteristics of the probe, such as internal energy, spatial
extension, and spin, is analyzed. 
 
The paper is organized as follows: the definition of the centroid in a
curved spacetime, as well as elementary facts about de Sitter spacetime, are 
discussed in section \ref{S2}. The calculation of the effective sectional
curvature, using three different probes, is carried out in section~\ref{S3}, which ends 
with a discussion of some aspects of our approach.
Finally, in section~\ref{S4}, further refinements
of our prescription are sketched out, and some remarks are offered on the
relevance of these preliminary results to the search for a theory of quantum
gravity.
%%%%%%%%%%%%%%%%%%%%%%%%%%%%%%%%%%%%%%%%%%%%%%%%%%%%%%%%%%%%%%
%%%%%%%%%%%%%%%%%%%%%%%%%%%%%%%%%%%%%%%%%%%%%%%%%%%%%%%%%%%%%%
%%%%%%%%%%%%%%%%%%%%%%%%%%%%%%%%%%%%%%%%%%%%%%%%%%%%%%%%%%%%%%
\section{Background}
\label{S2}
%%%%%%%%%%%%%%%%%%%%%%%%%%%%%%%%%%%%%%%%%%%%%%%%%%%%%%%%%%%%%%
\subsection{Centroid in a curved spacetime} 
\label{Ciacs}
%%%%%%%%%%%%%%%%%%%%%%%%%%%%%%%%%%%%%%%%%%%%%%%%%
As the main thesis advocated in this paper is served equally well by any
reasonable generalization of the Newtonian center-of-mass to curved backgrounds,
we choose, for convenience, to work with the \emph{centroid}. The composite
objects that we use as probes consist in a collection of free point particles
--- in this case, the \emph{special} relativistic definition \cite{Pry:48} of the
centroid \wrt{} an observer%
\footnote{%
As the last phrase implies, the centroid of an extended
object is, in general, an observer-dependent concept, as it depends both on his position and his velocity.}%
, involves an average of the (vector) positions of
the particles, weighted by their energies. 

The generalization of the above concept
to arbitrary backgrounds proceeds as follows.
Consider a spacetime $\calM$ with a metric $g$, an extended object $\calO$
in it and an observer $A$ with respect to whom the centroid of $\calO$ is to be
found. 
Call $W$ the observer's worldline, and, given a particular point $x$ on it, 
construct a hypersurface $\Sigma_x$, normal to $W$ at $x$, by extending all 
geodesics%
\footnote{%
We assume that the energy-momentum tensor of the probe has support in a region where the above geodesics 
do form a hypersurface --- this is guaranteed  if the particle's spatial
extension is, at all times, much smaller than the local radius of curvature of
$\calM$.}, orthogonal to $W$, emanating from $x$. $\Sigma_x$ plays the role of a 
simultaneity hypersurface for the observer $A$ at $x$, whose four-velocity we
denote by $u$. The worldlines of the
particles comprising $\calO$ intersect $\Sigma_x$ at the points $z_i$.
The vector ``position'' assigned to the $i$-th particle with respect
to the observer $A$ at $x$, with four-velocity $u$, is given by the
vector $\Xi _i$ in the tangent space at $x$, which is orthogonal to $u$, and
such that $\operatorname{exp}(\Xi_i)=z_{i}$. 
Moreover, the ``energy'' $E_{i}$ assigned to this particle is
obtained by parallel transporting the observer four-velocity $u$ to $z_i$ along the
(assumed) 
unique geodesic connecting $x$ and $z_i$, and subsequently projecting it onto
$p_i$, \ie, $E_i=-g(\tilde{u}, p_i)$, where
$\tilde{u}$ is the transported four-velocity. The position vector of the centroid of
$\calO$ \wrt{} $x$ and $u$, $\Xi(x,u,\calO)$, is then given by the sum of the
$\Xi_i$, weighted by the relative energies, 
\ble{XixvxO}
\Xi(x,u,\calO)= \frac{\sum_i E_i \, \Xi_i}{\sum_j E_j} \, , 
\ee
which is mapped, by the exponential map, to the centroid's position
$Z(x,u,\calO)$ on $\calM$, 
\ble{ZxvxO} Z(x,u,\calO)
=
\operatorname{exp}\left(\Xi(x,u,\calO)\right) 
\, . 
\ee
This construction can be repeated at every point of $W$, giving rise to
the centroid worldline $C(W,\calO)$. It is exactly this curve that we consider
as an effective geodesic of the spacetime $\calM$, observed by $A$ (with
worldline $W$), by using the probe $\calO$.
%%%%%%%%%%%%%%%%%%%%%%%%%%%%%%%%%%%%%%%%%%%%%%%%%%%%%%%%%%%%%%
%%%%%%%%%%%%%%%%%%%%%%%%%%%%%%%%%%%%%%%%%%%%%%%%%%%%%%%%%%%%%%
\subsection{De Sitter spacetime}
\label{DSs}
%%%%%%%%%%%%%%%%%%%%%%%%%%%%%%%%%%%%%%%%%%%%%%%%%%%%%%%%%%%%%%
We follow Weinberg's \cite{Wei:72} notation (with $K \rightarrow 1$), 
taking the metric for de Sitter spacetime to be
\ble{g1p1}
(g_{\mu \nu})
=
\left(
\begin{array}{ccc}
-1 & 0 & 0
\\
0 & 1 & 0
\\
0 & 0 & 1
\end{array}
\right)
+\frac{1}{1+t^2-x^2-y^2}
 \left(
\begin{array}{ccc}
t^2 & -t x & -t y
\\
-t x & x^2 & x y
\\
-t y & x y & y^2
\end{array}
\right)
\,  ,
\ee
where only the $(t,x,y)$ part is shown, the rest of the coordinates being
mere spectators in what follows. 
The corresponding affine connection is
\be
\Gamma^\mu_{\phantom{\mu}\nu \lambda} = x^\mu g_{\nu \lambda}
\, ,
\ee
implying that affinely parametrized geodesics $x(s)$ satisfy
\ble{geoeq}
\ddot{x}^\mu \pm x^\mu=0
\, ,
\ee
where the dot denotes derivative \wrt{} arclength $s$, and the plus (minus) sign
applies to spacelike (timelike) curves. 
It follows that the coordinates $x^\mu(s)$ of spacelike (timelike) geodesics are
linear combinations of circular (hyperbolic) cosines and sines of $s$,
with the coefficients constrained by the condition 
$g_{\mu \nu} \dot{x}^\mu \dot{x}^\nu=\pm 1$.
%%%%%%%%%%%%%%%%%%%%%%%%%%%%%%%%%%%%%%%%%%%%%%%%%%%%%%%%%%%%%%
\section{Towards an Operational Classical Geometry}
\label{S3}
%%%%%%%%%%%%%%%%%%%%%%%%%%%%%%%%%%%%%%%%%%%%%%%%%%%%%%%%%%%%%% 
We explore now de Sitter spacetime, using several classical extended
probes. We construct these out of two free point
particles of unit mass, and we arrange for their centroids to pass through the
origin. By varying the initial conditions we get, in each case, a family of
neighboring centroid worldlines, the relative acceleration of which is used to
define an effective sectional curvature in the $t$-$x$ plane at the origin. 

In all cases considered, the observer is at rest in the frame employed, with
worldline given by
\ble{ow}
t_{\text{obs}}(s)=\sinh s
\, ,
\qquad
x_{\text{obs}}(s)=0
\, ,
\qquad 
y_{\text{obs}}(s)=0
\, ,
\ee
where $s$ denotes proper-time, and four-velocity
\ble{obs4vel}
u=
\left(
\cosh s,0,0
\right)
=
\left(\sqrt{1+\tau^2},0,0\right)
\, ,
\ee
the latter form being in terms of $t_\text{obs}(s)\equiv \tau$.  
The observer's simultaneity surface $\Sigma$ at $\tau$, 
generated by geodesics orthogonal to $u$, is given by
\ble{Sigmacoord}
t_{\Sigma}(\lambda,\phi)=\tau \cos \lambda
\, ,
\qquad
x_{\Sigma}(\lambda,\phi)=c_\phi \sin \lambda
\, ,
\qquad 
y_{\Sigma}(\lambda,\phi)=s_\phi \sin \lambda
\, ,
\ee
where $c_\phi \equiv \cos(\phi)$, $s_\phi \equiv \sin(\phi)$.
Notice also that parallel transport of $u$ from $(\tau,0,0)$ to a general
point $P=(t_\Sigma,x_\Sigma,y_\Sigma)$ on $\Sigma$, along the (assumed unique)
geodesic connecting them, leaves its components unchanged.
Finally, the sectional curvature $K$ of the $t$-$x$ plane at the origin, for
the metric~(\ref{g1p1}), is equal to $-1$ --- our effective sectional curvature
$K_\text{eff}$ should tend to this value as our probes become pointlike.
%%%%%%%%%%%%%%%%%%%%%%%%%%%%%%%%%%%%%%%%%%%%%%%%%%%%%%%%%%%%%%
\subsection{Point hot probe}
\label{PhpI}    
\begin{figure}[t]
\centering
\psfig{figure=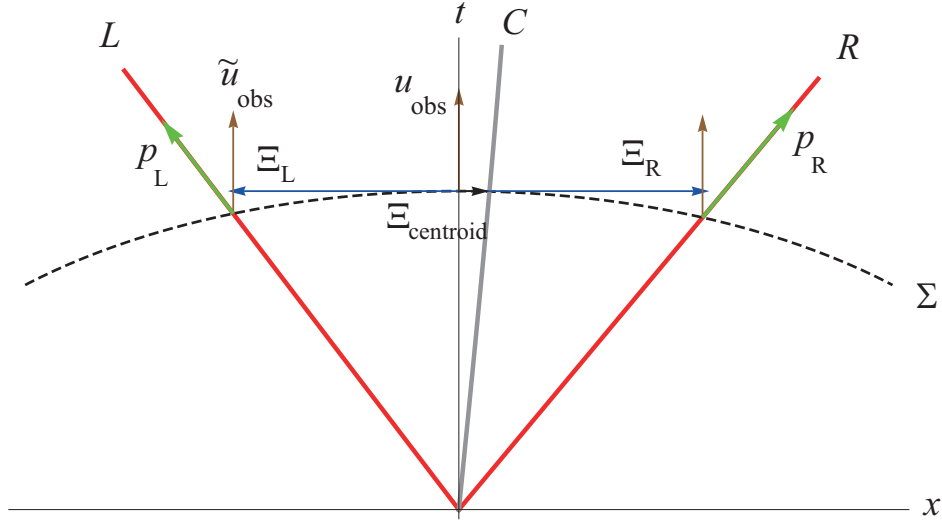,width=0.8\textwidth}
\caption{Setup for the calculation of the sectional curvature at the origin
using a point hot probe (see section \ref{PhpI}) --- the values of the
parameters used are $\epsilon=.2$, $\eta=1.0$, and $\tau=\sinh .5$.}
\label{fig:ppI}
\end{figure}

%%%%%%%%%%%%%%%%%%%%%%%%%%%%%%%%%%%%%%%%%%%%%%%%%%%%%%%%%%%%%%
The name of our first probe derives from the fact that, at $t=0$, it has zero
spatial extension but nonzero internal energy. 
It consists of two free point particles (L and R), moving in opposite directions
along the $x$-axis, with infinitesimally
different ``rapidities'' --- their worldlines are given by   
\begin{alignat}{3}
 \label{ppI_trajectories}
t_{\text{L}}(s) &= c_\eta \sinh s
\, ,
& \qquad
x_{\text{L}}(s) &= -s_\eta \sinh s
\, ,
\\
t_{\text{R}}(s) &=c_{\eta+\epsilon} \sinh s
\, ,
& \qquad
x_{\text{R}}(s) &= s_{\eta+\epsilon} \sinh s
\, ,
\end{alignat}
where $s$ is the proper-time of each particle, and $\epsilon \ll 1$, 
$s_\eta \equiv \sinh \eta$, $c_\eta \equiv \cosh \eta$, \etc{} (see
Fig.~\ref{fig:ppI}).
%%%%%%%%%%%%%%%%%%%%%%%%%%%%%%%%%%%%%
%%%%%%%%%%%%%%%%%%%%%%%%%%%%%%%%%%%%%

Consider now a free particle moving along the $x$-axis with generic rapidity
$\eta$, its worldline being
\ble{genwl}
t(s)=c_\eta \sinh s
\, ,
\qquad 
x(s) = s_\eta \sinh s
\, .
\ee
The above worldline crosses  $\Sigma$ at the event 
\ble{XTcoord}
(T,X)=\frac{\tau}{\sqrt{1+\beta^2}}(1,v)
\ee
(where $v \equiv \tanh \eta$ and $\beta \equiv v \tau$), which lies a geodesic
distance $S= \arcsin \beta/\sqrt{1+\beta^2}$ from the observer at
$t_{\text{obs}}=\tau$. The particle's position vector at
$t_{\text{obs}}=\tau$ is then
\ble{ppv}
\Xi(\eta) = (0,S)
\, ,
\ee
with $|\Xi|=\big(g_{\mu \nu}(\tau,0)\Xi^\mu \Xi^\nu\big)^{1/2}=S$ (note
that
$\Xi$
lives in the tangent space at  $(\tau,0)$).

The momentum of the particle, assuming unit mass, is given by
\ble{pdef}
p=\partial_s(t(s),x(s))= (c_\eta  \cosh s,s_\eta  \cosh s)
\, ,
\ee
so that the particle's energy comes out to be
\ble{energyppI}
E(\eta)=-g_{\mu \nu}(T,X)\tilde{u}^\mu p^\nu =
\frac{1+s_\eta^2\cosh^2 s}{c_\eta \cosh
s\sqrt{1-c_\eta^{-2}\tanh^2 s}}
 \, .
\ee
In terms of $\Xi(\eta)$, $E(\eta)$, the probe's centroid is given by
\ble{Xicmph}
\Xi_{\text{centroid}}
=
\frac{E(-\eta)\Xi(-\eta)+E(\eta+\epsilon)\Xi(\eta+\epsilon)}{
E(-\eta)+E(\eta+\epsilon)}
\, ,
\ee
the resulting expression being too long to include here --- notice
though that $\Xi_{\text{centroid}}=\calO(\epsilon)$. Finally, the exponential
map of $\Xi_{\text{centroid}}$ gives the centroid position --- the
corresponding worldline $C$,
obtained by varying $\tau$, is plotted in grey in figure~\ref{fig:ppI}. The
centroid moves slowly ($\sim \calO(\epsilon)$) to the right, as a result of
the slightly higher rapidity of the right particle. 

\begin{figure}[h]
\centering
\psfig{figure=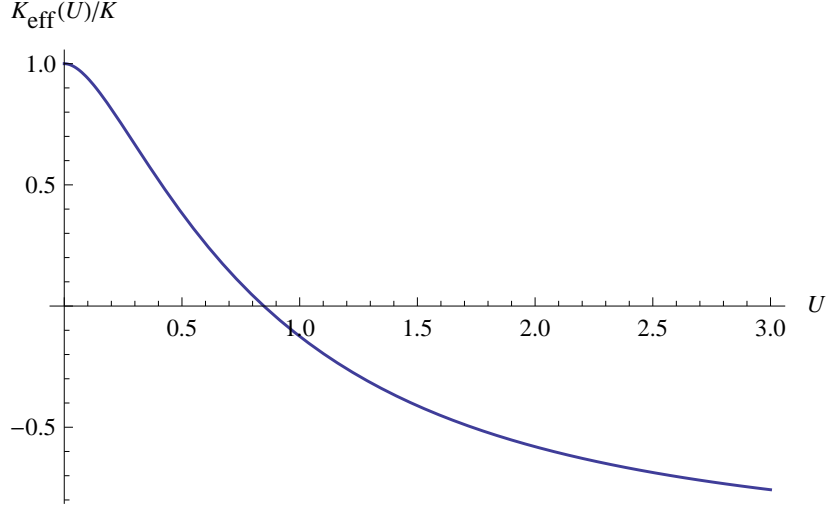,width=0.7\textwidth}
\caption{Plot of the effective curvature at the origin for a point hot probe, in
units of the background one $K=-1$, as a function of the initial internal energy
per constituent particle $U$.}
\label{fig:Keffeta}
\end{figure}

What we have computed so far is the trajectory of the effective position (centroid) of the freely moving extended probe. It is easily seen that $C$ is not, in general, a geodesic of the underlying de Sitter metric. However, our strategy is to let it play a similar role for the effective geometry we intend to extract. To this end, we study how neighboring such centroid worldlines accelerate with respect to each other and arrive, \emph{a la} Jacobi, to a concept of effective sectional curvature. To get a neighboring worldline we could change $\epsilon$ by a little bit, but this would amount to employing a manifestly different object, a choice that would lead to a poorly designed experiment: the separation vector of neighboring geodesics would get, in this case, a contribution from the $\epsilon$-dependence of the probe itself. The best option seems to be to consider two identical probes moving in two slightly different directions. For a
general spacetime, ``sameness'' of distinct objects is not an available
concept, but the symmetries of de Sitter spacetime allow us to overcome this
difficulty: all we do is reflect the setup around the $t$-axis and obtain a
second worldline $C'$ traveling to the left, the important point being that $C'$
describes the motion of the \emph{same} probe thrown with different initial
conditions. In the limit $\epsilon \rightarrow 0$, the separation vector between
$C$ and $C'$ is proportional to 
$J(s) = \partial_\epsilon \Xi_{\text{centroid}}|_{\epsilon=0}$ 
which, by symmetry, is orthogonal to the
observer four-velocity. A natural definition then for the $t$-$x$ sectional curvature at 
the origin, is%
\footnote{%
This is a simplified formula for the sectional curvature that is applicable in our case 
(see for example Ref. \cite{Fra:97}).}
\ble{Keffdef} 
K_{\text{eff}} 
\equiv 
\left. 
-\frac{\partial^2 |J(s)|/\partial s^2}{|J(s)|}
\right|_{s=0} 
\, ,
\ee 
where $\tau=\sinh s$ ought to be substituted in $J(s)$. Intermediate results are too 
lengthy to be included explicitly, but for $K_{\text{eff}}$ we do find a simple 
expression 
\ble{Keffresult} 
\frac{K_{\text{eff}}(\eta)}{K}
=
2\frac{\cosh 2\eta}{\cosh^4\eta}-1 
\approx 
1-2\eta^4+\frac{8}{3}\eta^6+\calO(\eta^7) 
\, , 
\ee
where $K=-1$ is the corresponding de Sitter sectional curvature. For $\eta$
tending to zero, the extended probe becomes point-like and $K_{\text{eff}}$
tends to $K$. Although our analysis is meant to hold primarily for small $\eta$,
so that the probe approximates reasonably a point particle, it is interesting to note
that for $\eta \approx 1.2$, $K_{\text{eff}}$ changes sign approaching
asymptotically, as $\eta$ tends to infinity, to $-K$. We may also express the
effective curvature as a function of the initial ($t=0$) internal energy per particle, 
$U \equiv \cosh \eta -1$, 
which may be regarded as a measure of the probe's initial temperature,
\ble{Keffresult1}
\frac{K_{\text{eff}}(U)}{K}
=
\frac{4}{(U+1)^2}-\frac{2}{(U+1)^4}-1
=
1-8 U^2+24 U^3+O\left(U^4\right)
\, .
\ee 
A plot of $K_{\text{eff}}(U)$ appears in
figure~\ref{fig:Keffeta}.
%%%%%%%%%%%%%%%%%%%%%%%%%%%%%%%%%%%%%%
%%%%%%%%%%%%%%%%%%%%%%%%%%%%%%%%%%%%%%
%%%%%%%%%%%%%%%%%%%%%%%%%%%%%%%%%%%%%%%%%%%%%%%%%%%%%%%%%%%%%%
\subsection{Extended cold probe}
\label{Ecp}
%%%%%%%%%%%%%%%%%%%%%%%%%%%%%%%%%%%%%%
\begin{figure}[t]
\centering
\psfig{figure=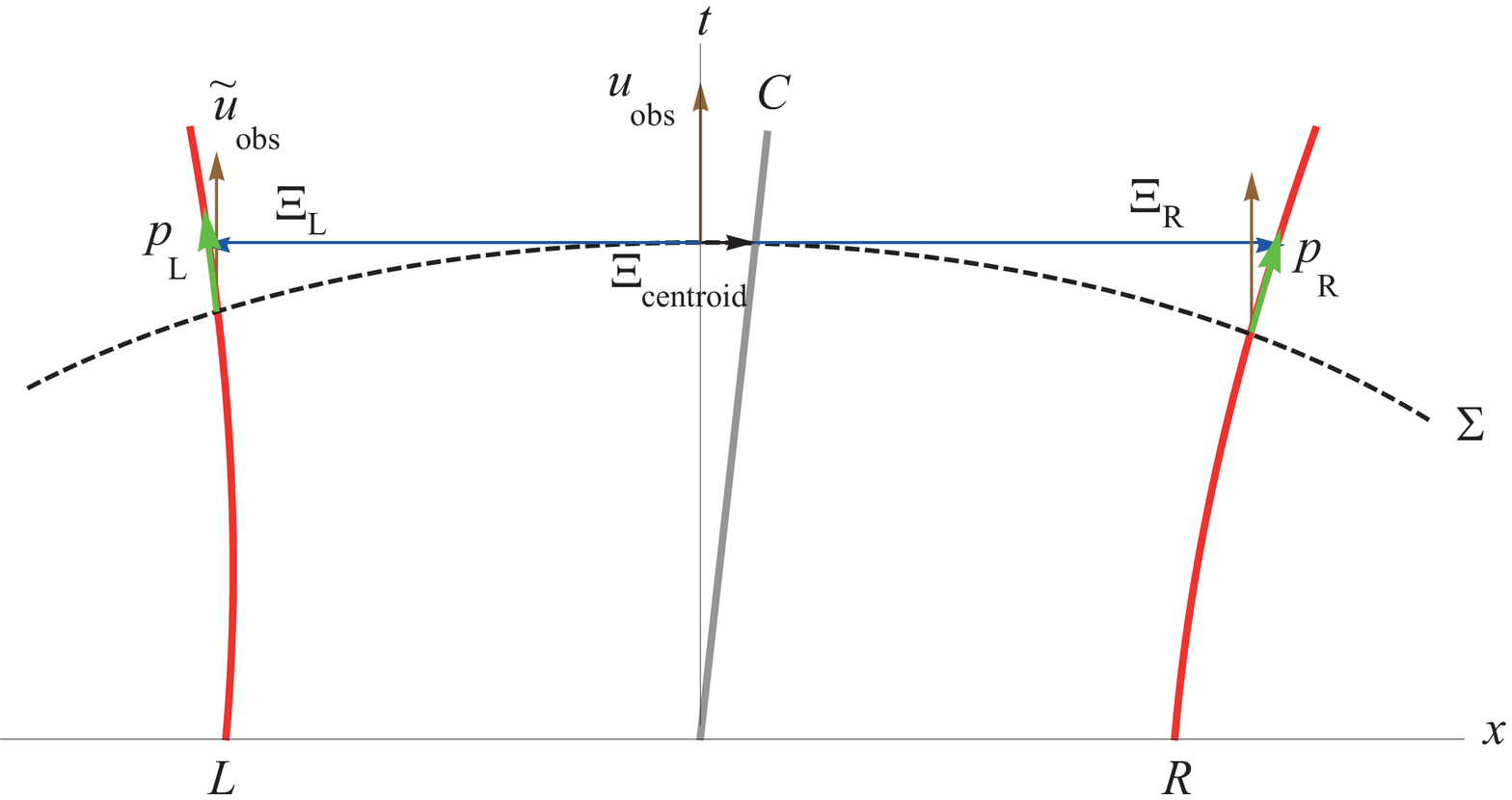,width=0.9\textwidth}
\caption{Setup for the calculation of the curvature at the origin using an
extended cold probe (see section \ref{Ecp}) --- the values of the parameters
used are $\epsilon=.1$, $a=.5$, and $t_0=\sinh .5$.}
\label{fig:CMa}
\end{figure}
\begin{figure}[t]
\centering
\psfig{figure=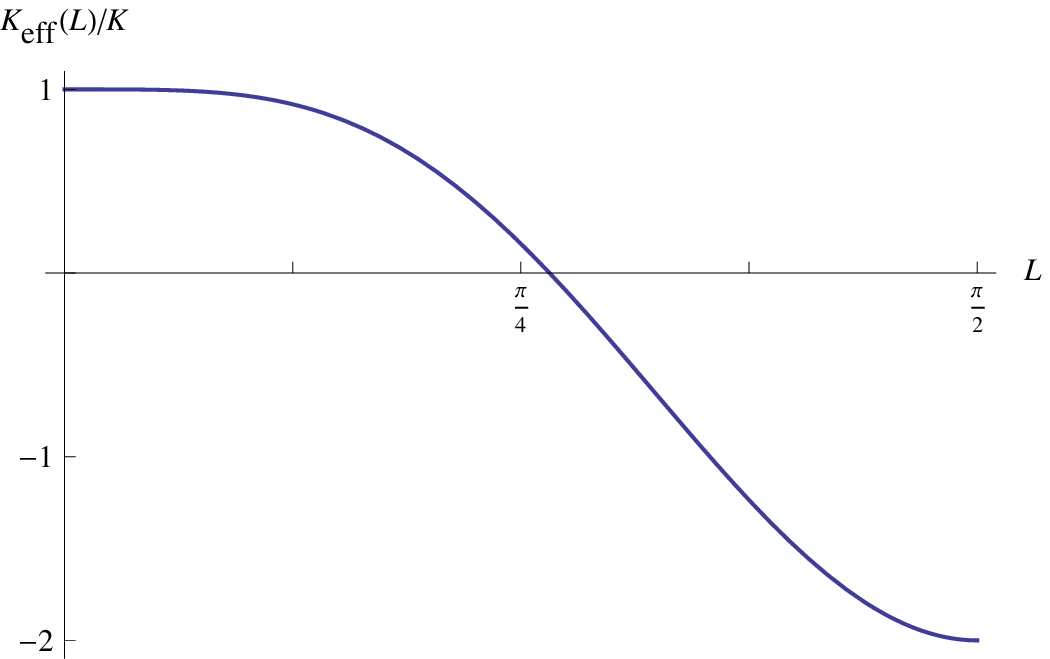,width=0.7\textwidth}
\caption{Plot of the effective curvature at the origin for the extended cold
probe, in units of $K=-1$, as a function of the initial half-length $L$ of the
probe (see section \ref{Ecp}).}
\label{fig:KeffL}
\end{figure}

 
%%%%%%%%%%%%%%%%%%%%%%%%%%%%%%%%%%%%%%
Our second probe is spatially extended, but contains essentially no internal energy, 
allowing us to isolate the
effect of its finite extension on the effective curvature
obtained by using it. The 
worldlines of the two free point particles (L and R) that comprise it are
\begin{eqnarray}
\label{ep_trajectories}
t_{\text{L}}(s)&= c_\epsilon \sinh s
\, ,
\qquad
x_{\text{L}}(s) &= -a \cosh s +b s_\epsilon \sinh s
\, ,
\\
t_{\text{R}}(s)&= c_\epsilon \sinh s
\, ,
\qquad 
x_{\text{R}}(s)&= a \cosh s +b s_\epsilon \sinh s
\, ,
\end{eqnarray}
($\epsilon \ll 1$, $b \equiv \sqrt{1-a^2}$, $s_\epsilon \equiv \sinh \epsilon$,
\etc{}), plotted in
red in figure~\ref{fig:CMa}. As in section~\ref{PhpI}, the problem is 1+1 dimensional, 
so that 
the calculation proceeds analogously to the previous case (intermediate results are too 
lengthy to be given here.) The effective curvature turns out to be
\ble{Keffresult2}
\frac{K_{\text{eff}}(a)}{K}
=\frac{1-4 a^2+3 a^4-2 b \left(3 a^2-2\right) a
\arcsin a}{1-a^2 +ab \arcsin a}
=1-4 a^4+\frac{8 a^6}{5}-\frac{16 a^8}{21}+O\left(a^9\right)
\, .
\ee
For $a$ tending to zero, the extended probe becomes point-like and
$K_{\text{eff}}$ tends again to $K$, while for
$a \approx 0.73$, $K_{\text{eff}}$ changes sign, approaching the
value $-2K$ as $a$ tends to 1 (limit in which the probe is of the same size
as the radius of the space being probed). $K_{\text{eff}}$ may also be expressed 
as a function
of the initial half-length $L\equiv \arcsin a$ of the probe,
\bae
\label{Keffresult3}
\frac{K_{\text{eff}}(L)}{K}
& = &
\frac{-2 L \sin L+6 L \sin 3 L+\cos L+3 \cos 3 L}{4 L \sin L+4\cos L}
\nonumber \\
& = &
1-4 L^4+\frac{64 L^6}{15}-\frac{332 L^8}{105}+O\left(L^9\right)
\, .
\eae
A plot of 
$K_{\text{eff}}(L)$ appears in figure~\ref{fig:KeffL}.
%%%%%%%%%%%%%%%%%%%%%%%%%%%%%%%
%%%%%%%%%%%%%%%%%%%%%%%%%%%%%%%
\subsection{Spinning probe}
\label{Sp}
\begin{figure}[t]
\centering
\psfig{figure=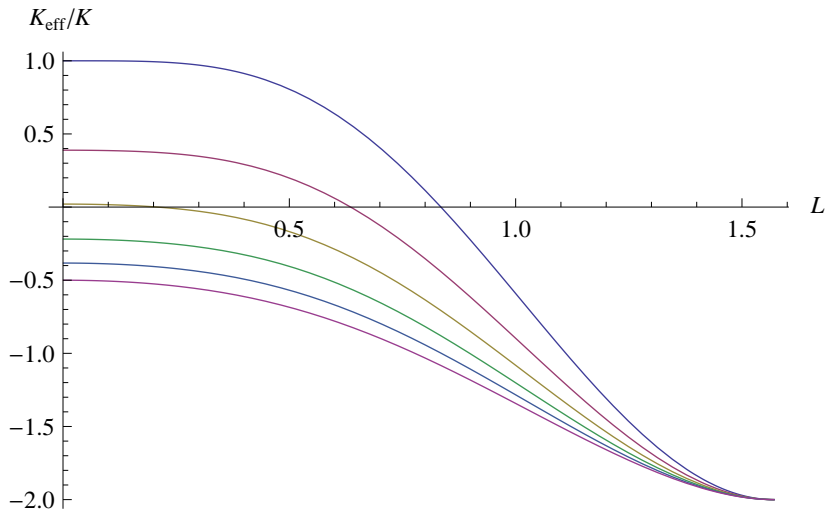,width=0.7\textwidth}
\caption{Plot of the effective curvature at the origin for the spinning probe,
in units of $K$, as a function of $L$, for
various values of $U$ (see section \ref{Sp}). The six curves shown correspond
to $U$ ranging from 0 (top) to 1 (bottom), in increments of .2.}
\label{fig:Keffaeta}
\end{figure}

%%%%%%%%%%%%%%%%%%%%%%%%%%%%%%%%%%%%%%%%%%%%%%%%%%%%%%%%%%%%%%
Finally, we study the effect of the probe's spin (internal angular momentum) on the 
effective curvature by considering the following worldlines for the constituent $L$ and 
$R$ particles,
\begin{alignat}{5}
\label{sp_trajectories}
t_{\text{L}}(s)
&= 
c_\epsilon c_\eta \sinh s
\, ,
& \quad
x_{\text{L}}(s) 
&= 
-a \cosh s +b s_\epsilon \sinh s
\, ,
& \quad 
y_{\text{L}}(s) 
&= 
- c_\epsilon s_\eta \sinh s
\, ,
\\ 
t_{\text{R}}(s)
&= 
c_\epsilon c_\eta \sinh s
\, ,
& \quad
x_{\text{R}}(s)
&= 
a \cosh s +b s_\epsilon \sinh s
\, ,
& \quad 
y_{\text{R}}(s)
&= 
c_\epsilon s_\eta \sinh s
\, .
\end{alignat}
The above setup is obtained from that of the 
extended probe by imparting an additional transverse rapidity $\pm \eta$ to the 
constituent particles, so that the resulting probe has spin along the $z$-axis (some minor 
modifications in the coefficients of the hyperbolic functions of $s$ are needed in order 
to ensure that $s$ is actually proper-time.) We calculate, as before, the sectional 
curvature in the $t$-$x$ plane at the origin\footnote{For this particular extended probe, 
this is the only non-vanishing component of the effective curvature at the origin.}, 
finding
\begin{eqnarray}
\label{Keffresult4}
\frac{K_{\text{eff}}(a,\eta)}{K}
&=&
\frac{b}{2 a^3\cosh^2\eta \left(a^2-a b
 L-1\right)} 
\left[
 \left(6 a^5+2a^3-6a\right) b\right.\nonumber\\
 && \left.+\left(12 a^6-7 a^4-9 a^2+6\right) L-\cosh 2\eta 
 \left(
 4 a^3b-6ab
 +\left(
 a^4-9 a^2+6
 \right) L \right)
 \right]\nonumber\\
 &=& 
\left(
 1-4 a^4+O\left(a^6\right)
\right)
+\eta ^2
 \left(-2-\frac{4 a^2}{5}+\frac{428 a^4}{105}+O\left(a^6\right)\right)
\nonumber\\
&&
\qquad
+\eta ^4
 \left(\frac{4}{3}+\frac{8 a^2}{15}-\frac{856
 a^4}{315}+O\left(a^6\right)\right)
+O\left(\eta^6\right)
\, ,
\end{eqnarray}
where $b \equiv \sqrt{1-a^2}$ and $L \equiv \arcsin a$, as before. In the limit
$\eta \rightarrow 0$ we recover the extended probe result of the previous
section, equation~(\ref{Keffresult2}). On the other hand, in the $a \rightarrow
0$ limit we do not get the point probe result of section \ref{PhpI}, simply
because in the present case the internal motion of the constituent particles is
orthogonal to the direction of motion of the centroid, and not parallel to it as in
the point hot probe case. For large enough values of $a$ and/or $\eta$,
$K_{\text{eff}}$ changes sign, approaching the value $-2K$ when $a=1$, $\eta=0$,
and $-K$ when $a=0$, $\eta\rightarrow \infty$. As our probe has internal energy,
spatial extension, \emph{and} spin, it is not as immediate to isolate the effect
of each of these characteristics on $K_{\text{eff}}$. The power series expansion
in equation (\ref{Keffresult4}) sheds some light on this matter: the first
parenthesis, of $\calO(\eta^0)$, coincides with the \rhs{} of equation
(\ref{Keffresult2}), as already mentioned above. Of the $\calO(\eta^2)$ terms in
the second parenthesis, the first one, $-2\eta^2$, is an internal energy effect
that differs from the corresponding (null) term in equation (\ref{Keffresult})
for the reason mentioned at the beginning of this paragraph. The next term,
$-4\eta^2 a^2/5$, is the first contribution of the spin $S\sim \eta a$ to
$K_{\text{eff}}$ and is quadratic in $S$, odd terms being forbidden on symmetry
grounds. Subsequent $\calO(\eta^2)$ terms involve products of $S^2$ with ever
increasing powers of $a$, with similar remarks applicable to the rest of the
expansion. The effective curvature may also be expressed in terms of the
half-length $L=\arcsin a$ and the internal energy $U=\cosh \eta -1$. Plots of
$K_{\text{eff}}$ \emph{vs} $L$, for various values of $U$, appear in
figure~\ref{fig:Keffaeta}.
%%%%%%%%%%%%%%%%%%%%%%%%%%%%%%%
\subsection{Discussion}
\label{Discussion}
%%%%%%%%%%%%%%%%%%%%%%%%%%%%%%%
There are a few remarks that are appropriate at this point. To begin with, there is an 
apparent internal contradiction in our treatment, in that the centroid worldlines are 
used to define, essentially, 
an effective geometry, but then the given de Sitter background is employed, rather 
than this effective 
geometry, to quantify their relative acceleration. There are at least two ways to remedy 
this. On the one hand, one may invoke a perturbative approach to the problem, the 
perturbation parameter $\epsilon$ being, \eg, the ratio of the probe's size to the typical 
radius of curvature of the background spacetime. Then, to zeroth order in $\epsilon$, the 
effective geometry coincides with the background one. In such an approach, the difference 
between using the background or the effective geometry would only show up at a higher 
order in $\epsilon$, and our calculations are consistent, if truncated to the first 
nonvanishing $\epsilon$-correction. A second, related point of view, 
would require that the 
effective geometry coincide with the one used to determine it. In other words, an 
effective metric, or, perhaps, a more general geometrical entity (see remark below),  
are being sought, with the property that when these are used to quantify the relative 
acceleration of neighboring centroid worldlines, this same geometrical data is recovered. 
In this case, our calculations, without any truncation, are 
the first step in an iterative process, that would converge, if it converges at all, to 
the sought after effective geometry.

A second remark, already alluded to above, concerns the nature of the effective geometry 
that one may hope to arrive at. For example, one may ask whether  an 
effective metric exists, with respect to which the centroid worldlines are true 
geodesics.
General arguments suggest that this cannot be the case in general. The effective curvature 
we have computed above is but one of the many components of the effective Riemann tensor 
(assuming, for the moment, that the latter exists). To measure other components, probes 
have to be 
thrown along different coordinate axes, and there is no canonical way, in general, to 
guarantee that all probes needed to recover the full Riemann tensor are identical among 
themselves --- their extended nature renders such identification problematic. (An analogous problem arises when dealing with an effective connection, see Ref. \cite{Bon:12}.)
Since the effective 
Riemann tensor depends also on the probes, the obstruction 
to its existence is conceptual, rather than technical. It seems then that, in 
general, one may have to give up the hope for an effective geometry that mimics faithfully 
the standard one, although, in particular cases, with sufficient symmetry present, this 
might still be an attainable concept. When no effective metric can consistently be 
defined, a different set of geometrical data, that somehow incorporates probe information, 
may have to take its place.

A third remark, about probe design, is also due. The reader might feel that our probes
would be 
improved if some sort of rigidity was imparted to them, by connecting, for example, the 
constituent particles with elastic bands, or letting them interact electromagnetically. 
The way they are constructed above, out of free particles, they hardly conform to the 
standard preconception of a probe, as they spread out indefinitely with time. We can only 
agree with this criticism, but any attempt at ``gluing together'' the constituent 
particles ought to be accompanied by the inclusion of the energy-momentum tensor 
of the ``glue'' in the 
calculation of the centroid, a requirement that renders calculations intractable. On the 
other hand, better probes would of course probe better, but we doubt that the essence of 
any of our conclusions would be affected by using them.

Finally, we offer some thoughts on the steps that remain to be taken on the way to a satisfactory treatment of quantum probes. We focused so far on some of the effects of the probe's classical attributes on the corresponding effective geometry. When the probe is endowed with true quantum nature, it is natural to expect that the perceived geometry will also acquire quantum characteristics. Thus, just as the quantum probe explores, \emph{a la} Feynman, paths around the classical one, each weighted with a certain quantum amplitude, the corresponding geometrical quantities being measured might also exist in a superposition of classical states, their proper description being through wavefunctions, rather than definite numerical values as above. Another conceptual hurdle might emerge due to the uncertainty principle. Notice, for example, that we have computed above the effective sectional curvature at a point of spacetime as a function of the probe's energy. In a quantum treatment, we expect a tension between the 
necessity of localizing the probe, so that we can talk of the geometry \emph{at a point}, with the resulting spread of the probe's momentum and energy, as dictated by Heisenberg's principle --- similar inconsistencies lurk, for example, in ``rainbow gravity'' proposals, where the effective metric perceived by a probe at a point is assumed to be a function of the probe's energy. Other subtleties may be encountered in doing quantum mechanics on curved spacetimes. Apart from the well known ordering ambiguities, and the possible breakdown of the test particle assumption, leading, in extreme cases, to black hole formation \cite{BlackHoleFormation,Nee:94,Gam_Por_Pul:04}, we also expect inconsistencies with the single particle picture inherent in our analysis, as the relativistic treatment necessary in this case allows particle creation and annihilation, making eventually inevitable a quantum field theoretical approach. 

Keeping these considerations in mind, one may try to generalize our results in a sequence of steps. First, it is straightforward, at least conceptually, to describe probes with continuous matter distribution. Then, quantum probes may be introduced, and conditions determined for the single particle approximation to be valid. In this restricted setting, an effective geometry could be determined, that might exhibit quantum features, as alluded to above. The quantum states considered for the probes ought to resemble coherent states, so that the emerging quantum geometry could be thought of as some sort of deformation of our present results. The consideration of more general quantum states would take us beyond the paradigm of classical geometry, deep into unchartered territory. Finally, relaxing the single particle conditions, thus embracing the full complexity of quantum fields, would radically alter the nature of our inquiries --- probing geometry in this regime would bear resemblance to the old conundrum of 
determining the shape of a drum by the sounds it produces.  
%%%%%%%%%%%%%%%%%%%%%%%%%%%%%%%%%%%%%%%%%%%%%%%%%%%%%%%%%%%%%%
%%%%%%%%%%%%%%%%%%%%%%%%%%%%%%%%%%%%%%%%%%%%%%%%%%%%%%%%%%%%%%
\section{Summary and Conclusions}
\label{S4}
%%%%%%%%%%%%%%%%%%%%%%%%%%%%%%%%%%%%%%%%%%%%%%%%%%%%%%%%%%%%%%
%%%%%%%%%%%%%%%%%%%%%%%%%%%%%%%%%%%%%%%%%%%%%%%%%%%%%%%%%%%%%%
We studied the effects produced by the internal energy, finite extension, and
spin of classical probes when measuring the sectional curvature in de Sitter
spacetime. The results found are summarized in figures \ref{fig:Keffeta},
\ref{fig:KeffL}, and \ref{fig:Keffaeta}. For small values of the parameters, the
absolute value of the effective curvature $K_{\text{eff}}$ is, generically, diminished,
in comparison to the de Sitter value, while for very energetic or sufficiently large
probes, a change of sign in $K_{\text{eff}}$ occurs. 
In all cases, as our probes tend to point particles, the sectional curvature of de
Sitter spacetime is recovered. The fact that the deviations from the 
background geometry are
small when the extended particles are nearly point like, fits well with the
agreement of the general relativistic predictions with observations, which are
done with real objects.

The results found naturally depend on the probes, making the effective geometry advocated 
here a relational one, where the totality of the system involved, consisting of the
manifold under study, the probe, 
and the particular experiment employed, form a tightly interwoven whole, to which the 
traditional notion of geometry is only an approximation. The 
conclusion is then reached that, if the probes considered are quantum, the geometry of 
classical manifolds cannot, even in principle, be operationally defined in a probe 
independent fashion. The conceptual core of these findings is expected to carry over to 
the quantum gravity case, casting doubts on attempts to extract a classical geometry,
as an approximate description of the underlying quantum state of the gravitational degrees 
of freedom, while failing to incorporate explicitly matter and particular 
experiments
\footnote{
For complementary discussions of the relational point of view advocated here see, \eg, \cite{Rov:96,relational,Bar:10}.}. Clearly, the problem contemplated here is 
nontrivial and its full solution should have repercussions  in the evaluation of 
candidates for a quantum theory of
gravity\footnote{There are some superficial similarities between the present work and  
studies in 
cosmology, where local inhomogeneities are taken as modifying the effective average 
geometry. It is well know that in the Raychaudhuri equations, describing the behavior of 
the expansion of geodesic congruences, the twist and shear can be interpreted as effective 
contributions to the energy-momentum of matter fields \cite{Cosmology-Raychaudhuri}.}.
 
We reiterate that the prescription we follow in this work is a hybrid between
the textbook idealizations of classical geometry and the fully operational
geometry we advocate, in that it uses the underlying metric, \eg, to quantify
the relative acceleration of neighboring centroid worldlines. In a truly
operational treatment, distances would be measured in terms of
``standard rods'', or, better, light signals and clocks, the latter realized by 
particular oscillators, the quantum nature of which would prevent arbitrary 
precision (related matters have been studied in, \eg, \cite{Wig:57,Salecker,Aha_et_al:98,Ame:94,Ng_vDam:95}. Our intention here
is slightly less ambitious in that we focus on \emph{some} of the effects
produced by realistic probes due to their various physical characteristics,
assuming, for simplicity's sake, that background metric information is given.
 Whatever the fate of these further explorations, the main lesson
to be drawn from the present work is the relational nature of quantum geometry,
that most of the mainstream approaches seem to ignore.
%%%%%%%%%%%%%%%%%%%%%%%%%%%%%%%%%%%%%%%%%%%%%%%%%%%%%%%%%%%%%%
\section*{Acknowledgments}

The authors would like to thank J.D. Bekenstein for pointing out 
reference \cite{Salecker}. Partial financial support from the UNAM-DGAPA projects
IN 114712, IN 118309 (PA, CC), IN 107412 (YB, DS), and CONACyT projects 103486 (PA, YB, 
CC),
101712 (YB, DS) is gratefully acknowledged. 

%\bibliographystyle{plain}
%\bibliography{strings}
%%%%%%%%%%%%%%%%%%%%%%%%%%%%%%%%%%%%%%%%%%%%%%%%%%%%%%%%%%%%%%

\end{document}